\documentclass[sigconf, 10pt, authorversion,nonacm]{acmart}
\newcommand{\fakeparagraph}[1]{\vspace{.5mm}\noindent\textbf{#1.}}
\newcommand{\fakepar}[1]{\fakeparagraph{#1}}
\newenvironment{itemize*}{\begin{itemize}}{\end{itemize}}

\usepackage{csquotes}
\usepackage{pgfplots}
\usepackage{adjustbox}
\usepackage{mdframed}
\AtBeginDocument{%
  }

\begin{document}

\title[VoCopilot: Voice-Activated Tracking of Everyday Interactions]{VoCopilot: Voice-Activated Tracking of Everyday Interactions}

\author{Sheen An Goh}
\affiliation{
  \institution{National University of Singapore}
  \country{}
}
\email{e0926997@u.nus.edu}

\author{Manoj Gulati}
\affiliation{
  \institution{National University of Singapore}
  \country{}
}
\email{manojg@nus.edu.sg}

\author{Ambuj Varshney}
\affiliation{
  \institution{National University of Singapore}
  \country{}
}
\email{ambujv@nus.edu.sg}

\begin{abstract}
Voice plays an important role in our lives by facilitating communication, conveying emotions, and indicating health. Therefore, tracking vocal interactions can provide valuable insight into many aspects of our lives. This paper presents our ongoing efforts to design a new vocal tracking system we call VoCopilot.  VoCopilot is an end-to-end system centered around an energy-efficient acoustic hardware and firmware combined with advanced machine learning models. As a result, VoCopilot is able to continuously track conversations, record them, transcribe them, and then extract useful insights from them. By utilizing large language models, VoCopilot ensures the user can extract useful insights from recorded interactions without having to learn complex machine learning techniques. In order to protect the privacy of end users, VoCopilot uses a novel wake-up mechanism that only records conversations of end users. Additionally, all the rest of pipeline can be run on a commodity computer (Mac Mini M2). In this  work, we show the effectiveness of VoCopilot in real-world environment for two use cases.
\end{abstract}

\maketitle

\section{Introduction}
Wearable devices have witnessed a significant interest. They are enabled by the confluence of advances in the sensor miniaturization, computing capabilities, and reduced power consumption of  electronic circuits. Today, embedded and mobile devices monitor our steps~\cite{brajdic2013walk}, sleeping~\cite{sleepdetection}, hand washing~\cite{handwash} and many other activities. Wearable devices come in different shape and size, with smartwatches being common, and rapid growth in earphones~\cite{earable} and rings as a sensing platform. However, they miss out on tracking the most fundamental of human activity, our vocal interactions.

Vocal interactions serve as a window into our feelings, thoughts, and well-being~\cite{han2022sounds, heartsensing}.  Today, continuous tracking and extraction of insights from our vocal interactions remains a  challenge. While tools like voice recorders, smartwatches, earphones, and smartphones, exist, none of them are suitable for continuous tracking of vocal interactions.

A smartphone can track our vocal interactions. However, smartphones are general purpose devices, and  re-purposing them for vocal tracking leads to rapid depletion of its battery~\cite{acousticsensing}, also negatively impacting its core functionality.  Moreover, a smartphones microphone is hardly in an optimal position for day-long vocal tracking because the phone is typically stored in pockets or bags. Smartwatches and earphones suffer from similar challenges, including the shorter battery lifespan. The closest device  for vocal tracking are voice recorders. They record extended interactions, but their bulky design, short battery lifespan, and lack of features, such as detecting emotions or transcribing voice, limits their usefulness. Furthermore, these devices  pose significant privacy concerns since they commonly record interactions without an explicit consent  of all the involved participants. 

\begin{figure}[!tb]
 \centering
   \includegraphics[width=\columnwidth]{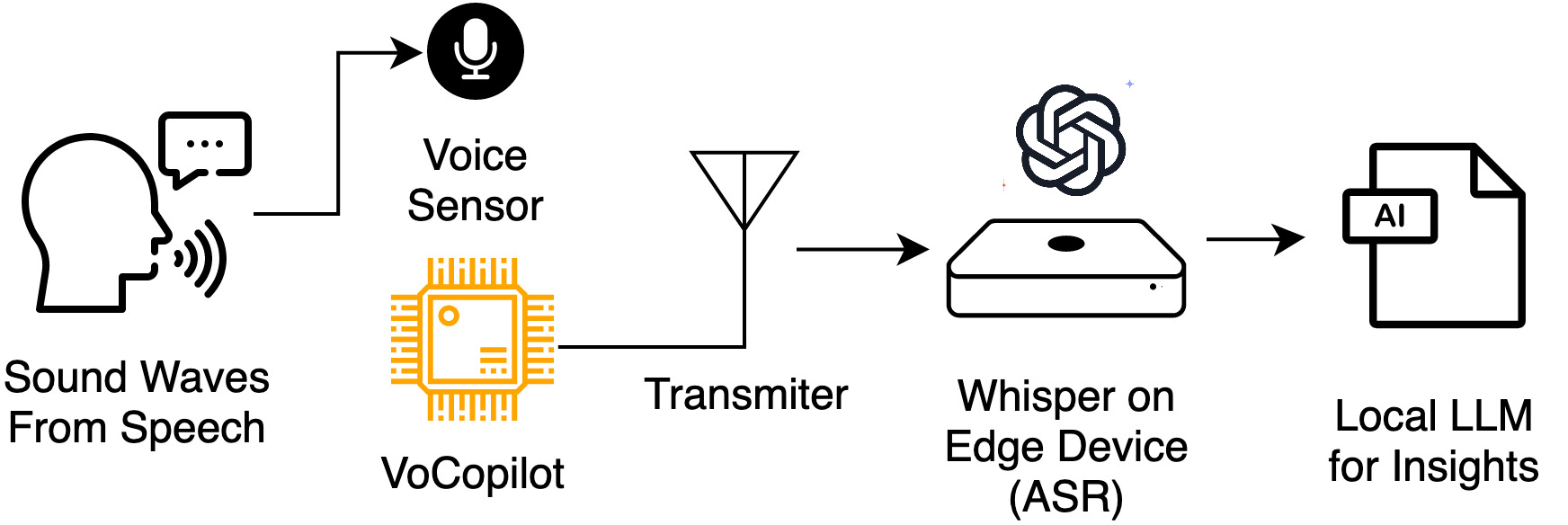}
      \vspace{-6mm}
   \caption{VoCopilot continuously tracks interactions for keywords, and only records when specific keywords are heard. The recorded interactions are sent to an edge device where they are transcribed using an automatic speech recognition system~(OpenAI Whisper). On the basis of the user's inputs provided as a prompt in the natural language, a local language models analyses and extract insights from transcribed interactions.}~\label{fig:overview}
   \vspace{-8mm}
   \label{overview}
 \end{figure}

In this work, we tackle the challenge of continuous tracking of the vocal interactions. We present a design of an end-to-end system that we call VoCopilot. This system can capture interactions, and analyse them for applications such as transcribing conversations, tracking emotions and others. We provide an overview of the system in the Figure~\ref{overview}.

\fakepar{Design} VoCopilot is designed to ensure low-power consumption of the tracker. This can ensure that the tracker can be instantiated in form-factor of weable devices. No

Continuous tracking of acoustic emissions is an energy expensive process. Hence, we have designed VoCo

have designed the system to ensure the low-power consumption for continuous tracking, and to ensure privacy while capturing and analysing the recorded conversations. In particular, we anticipate the VoCopilot to be designed as a tracker that could be worn by the user throughout the day. The tracker could be designed to have a small form factor like a pendant. The recorded interactions could then be analysed by the user by providing natural language prompts to extract meaningful insights from the day. Hence,  the system has two components: a tracker, and an edge device that analyzes captured conversations.

An embedded system designed for acoustic sensing generally uses a pipelined architecture. Initially, a microphone picks up the sound waves. These are then amplified and converted to digital signals, which are processed by a microcontroller. Due to the significant energy consumption and complexity of onboard computation, this processing typically occurs offline at the device's edge. The traditional method, while effective, is energy-draining and captures indiscriminately all audible signals around it.

\begin{figure}[t]
\centering
\includegraphics[width=\columnwidth]{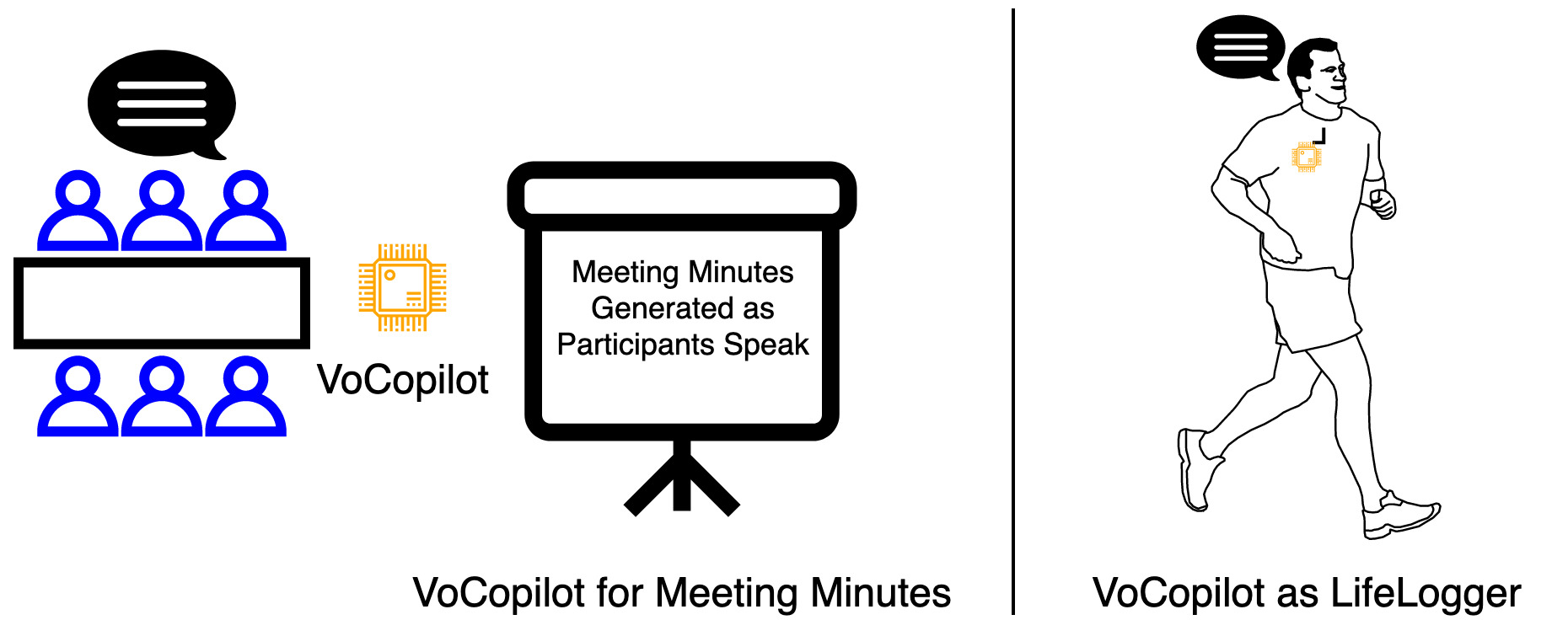}
\caption{\emph{Application scenarios.} VoCopilot can transcribe and give a summary of meetings. VoCopilot can also act as LifeLogger to give daily insights.}
\label{fig:meeting}
\vspace{-4mm}
\label{usecases}
\vspace{-4mm}
\end{figure}

Our first contribution is to design a novel audio capture mechanism that takes advantage of the unique capabilities of specialized accelerator chipsets. With just a hundred microwatts of power, they can track specific acoustic patterns. Thus, ensuring continuous tracking of the acoustic emissions. We build on this mechanism to design an energy-efficient tracker that only responds to specific keywords spoken by the end user. It serves as a mechanism to activate the energy-intensive recording process. This also ensures privacy, as it prevents recording of speech of other users.
Building on this mechanism, we design a tracker using the chipset NDP120 \cite{Syntiant_NDP120}, and demonstrate ability to track specific keywords with high accuracy. We also demonstrate ability to record conversations being triggered by the specific keywords.

Once the interactions are captured, they need to be analysed. Taking advantage of the emerging potential of transcriptional models, we have developed a mechanism that efficiently converts the recorded interactions into text. Specifically, we employ the  Whisper model by OpenAI~\cite{jacoby2021whisper}, which allows us to support multilingual transcription capabilities. 

The transcribed text need to be analyzed differently depending on the use case. Additionally, analyzing transcribed text requires expertise from the end user, making it difficult for them to use the system. In order to address this challenge, we employ a local language model (LLM) at the edge device. In this way, the user can provide prompts in natural language and instruct the language model to differently process
the text. Due to the fact that the model runs locally on the edge device, the conversation is also kept private, thus enforcing privacy of VoCopilot.

\fakepar{Use case} VoCopilot can enable several application scenarios. In this early work, we highlight two application scenarios. Firstly, we envision that VoCopilot could be used in meetings for recording and transcription. The device could be worn by participants or centrally positioned in the room. The system could record the ongoing conversation when it detects a specific word. Secondly, we see VoCopilot acting as a personal life-logger, monitoring a user's emotional state by identifying specific keywords, like terms indicating anger or tranquillity, or recording entire dialogues to create a real-time journal. We depict them in the Figure~\ref{usecases}.

\section{Related Work}
\fakepar{Low-power sensing} Designing energy-efficient acoustic systems has been an active area of research. Talla et al.~\cite{bfcellphone} demonstrate a low-power platform that tracks speech and transmit it using backscatter mechanism. Due to its low power consumption, it can be operated using energy harvested from photodiodes. Arora et al.~\cite{saturn} present a triboelectric-based low-power acoustic sensing system. Whisper proposed a novel back-scatter enabled acoustic sensing tag, powered using energy harvested from UHF RFID \cite{jacoby2021whisper}, which performs acoustic sensing with a low-power wake-on-audio microphone, discarding  noise using analog pre-gating stage. These systems, however, do not perform any processing on the device. Their goal is to perform energy-efficient acoustic sensing, and to transmit any sounds they sense. Moreover, they do not address the challenge of capturing information from non-consenting users.  Instead, we have developed an end-to-end system to ensure voice tracking is done in a low-power and privacy-preserving manner. In fact, VoCopilot can build on these advances, and would be our future-work.

\fakepar{Earphones and smartphones} Georgiev et. al. repurposed DSP core present on smartphones to enable continuous audio sensing~\cite{georgiev2014dsp}, and in a follow-up work, they enabled active listening tasks while optimising the energy footprint~\cite{georgiev2017low}. Nonetheless, their techniques are applicable to platforms with limited batteries, which face microphone placement challenges. Furthermore, we integrate language models into our end-to-end system to enable transcription and analysis of conversations beyond efficient acoustic sensing.

\fakepar{Language models and embedded sensing} The capabilities of language models have also advanced in recent years. There are claims that these models demonstrate sparks of artificial general intelligence~\cite{bubeck2023sparks}. There has been interest in leveraging language models with embedded and mobile systems~\cite{englhardt2023exploring}. Our approach combines advances in low-power acoustic sensing with language models to enable efficient transcription and analysis of vocal interactions.

%
%
%
%
%

\section{Design}
The VoCopilot system works as follows: The tracker device constantly listens for predetermined  key words in interactions. These keywords can activate ongoing  recordings of the interactions or indicate the speaker's emotional state. Thanks to a neural chipset (NDP120), the tracker constantly listens for keywords, while consuming microwatts of power. When the keywords are detected, the device activates its recording mechanism, archiving the interaction. Based on memory constraints, recording stops after a predetermined length of the interaction has been reached. An edge device receives recordings of the interaction via a wireless or direct connection. The edge device then uses a transcription system to translate the audio into text. The end-user provides prompts detailing the insights to be extracted, and an LLM extracts them from the archived recordings. Therefore, providing a comprehensive overview of the vocal interactions.

\fakepar{Tracker} We design the tracker to ensure low-power consumption to ensure prolonged operation on batteries. Hence, we design the tracker with a low-power audio wake-up mechanism, which is also one of the key contributions of this work. We design the wake-up mechanism using an energy efficient neural decision processor NDP120 from Syntiant~\cite{Syntiant_NDP120}. This processor is trained on voice of the end-user to detect specific keywords, which can be used to trigger the recording of the conversation. By capturing these words throughout the day, we could train the processor to detect words that indicate emotions, and thus enable the life-logging use case.

Once the specific keywords are detected, it wakes up the remainder of the platform which is then tasked with recording of the conversation. In this early work, we design the embedded platform using a low-power Cortex M4 ARM processor - Nordic NRF nRF52832. The entire platform is in a tiny form factor~(22.86 x 22.86 mm). To record conversations the platform uses low-power microphone Infineon IM69D130 interfaced through the PDM interface. The platform supports megabytes of onboard memory, nonetheless, to enable capturing of the longer conversations, we also interface an external micro-sd card, over SPI interface to the platform. 

\fakepar{Transcribing} Once the tracker has recorded interactions and communicated them to the edge device, the next step is to transcribe recorded interactions.  In particular, we use automatic speech recognition~(ASR) system. ASR have a long history dating  since 1980, employing methodologies like Hidden Markov models~(HMMs) and Gaussian mixture models~(GMMs). More recently, recent ASR models have used  Deep neural networks~(DNNs). Nonetheless, these models still suffered from challenges with achieving high accuracy. 

Transformer-based models have demonstrated significant improvements in capabilities over the last few years. This has also resulted in highly accurate transcriptions of conversations. Many of these models can even be run locally on the edge device. In this particular work, we employ one such model Open AI Whisper~\cite{whisper} to transcribe the recorded conversations. Based on the results presented in the work, it can transcribe in a reasonable amount of time. This model is freely available and open-source. The privacy of recorded interactions is also ensured by running locally.  As the edge device, we use an Apple Mac Mini M2 (16 GB RAM).

%
%
%
%
%

\fakepar{Extracting insights using language models} After transcription, these conversations are analyzed to provide a snapshot of the interactions. This could include summarizing a meeting or pinpointing crucial points of a conversation. As a result, the actions taken will depend on the user's specific needs. Traditionally, extracting such insights required an understanding of machine learning and natural language processing on the user's part. This is a complexity that might hinder the applicability of our system. Our solution circumvents this obstacle by leveraging a language model that runs locally on the edge device. The local language model also ensures the privacy of recorded interactions, since all the steps from transcription to analysis are performed locally.

We adopted Llama 2 \cite{touvron2023llama}, a collection of pretrained and fine-tuned LLMs ranging in a scale from 7 billion to 70 billion parameters. On most benchmarks, Llama 2 models outperformed open-source chat models, and they may be a suitable replacement for closed-source models like GPT-4. We evaluate its suitability for the edge device and present some results in the evaluation section. In particular, the user could provide appropriate prompts to extract the insights.

In tracking a meeting, we provide one such prompt and the LLM's response. Here is an example of a  meeting summary generated using VoCopilot leveragign the LLM.


\begin{mdframed}[backgroundcolor=gray!20]
  \textbf{RESPONSE (LLM):} Here are the three most important points: 1. The event, formerly known as eGovernment Global Exchange, has evolved into Digital Government Exchange (DGX). 2. DGX is held every two years and brings together leading practitioners from around the world to exchange ideas, tackle challenges, and foster an international community. 3. The event aims to provide a regular platform for those involved in making digital government a way of life to gather and share insights, with the goal of better serving citizens and businesses in a rapidly changing digital world.
\end{mdframed}

\section{Evaluation}
\pgfplotsset{compat=1.13}
\definecolor{findOptimalPartition}{HTML}{D7191C}
\definecolor{storeClusterComponent}{HTML}{FDAE61}
\definecolor{dbscan}{HTML}{ABDDA4}
\definecolor{constructCluster}{HTML}{2B83BA}

We conduct experiments to evaluate VoCopilot. The main findings of the results presented in this section are as follows:

\begin{itemize*}
    \item  Recorded conversations are small for onboard storage
    \item  Speaker and keywords identified with high accuracy 
    \item  End-to-end system achieves  highly accurate transcribing, and  analysis of the recorded interactions
\end{itemize*}

\fakepar{Memory storage requirements} The ability to capture conversations over an extended period can be helpful for applications such as taking meeting notes. This experiment examines the size of the audio file for recorded conversations. In order to keep the file size to a minimum, the audio files are encoded using the G722 codec. The conversations are recorded on an external SD card when the file size exceeds the platform's internal memory. We speak for different duration (14 times), and investigate the size of file with recorded conversation. We show the results in the Table~\ref{storage_table}.

As a result of the compression used, our results indicate that the file size is small. Additionally, the file size may be small enough to not require an external SD card for application such as life-logger, since we can typically store tens of minutes of conversations for just a few megabytes. These are typically the storage capacities available on commodity embedded platforms. Additionally, we can store conversations equivalent to 92.60 days by using a 64 GB SD Card and taking into account our storage usage of 8.000 KB per second for audio recording.




\begin{table}[ht]
  \vspace{-4mm}
  \centering
  \begin{adjustbox}{width=0.35\textwidth}
  \begin{tabular}{|c|c|c|}
  \hline
  \textbf{File Samples} & \textbf{File Size (KB)} & \textbf{Length (s)} \\
  \hline
  1 & 293 & 36.575 \\
  2 & 474 & 59.233 \\
  .. & .. & .. \\
  13 & 275 & 34.32 \\
  14 & 1144 & 143.018 \\
  \hline
  \textbf{Total} & 5911 & 738.912 \\
  \textbf{Average Size KB / s} & & 8.000 \\
  \hline
  \end{tabular}
\end{adjustbox}
  \caption{Storage required to store G722 Codec audio}
  \label{storage_table}
  \vspace{-10mm}
  \end{table}

\fakepar{Spotting keywords at a distance} Many of the use cases requires VoCopilot to be activated within the context of meetings, interaction with others or life logging. Typically, they require the tracker to be located at varied distances. Therefore, it is crucial for the tracker to exhibit high accuracy in keyword spotting~(KWS) over short to medium distances. Hence, we conducted an experiment to investigate the accuracy of KWS. This capabilities leverages the abilities of the NDP120.  As for methodology, we recorded the ratio of true positives per instance of the keywords spoken "Yes", "Stop", and "Go". This was carried out over varying distances, specifically with tracker located at distances of 10cm, 50cm, 1m, and 2m, which are typical distance that we can expect in the target  scenarios. We conduct two instance of experiment, with the keyword spoken 30 times in each run.

\begin{figure}[ht]
  \centering
\begin{tikzpicture}[draw, xscale=0.55, yscale=0.55]
\begin{axis}[
    title={KWS accuracy with the distance of tracker},
    xlabel={Distance},
    ylabel={Accuracy (\%)},
    xmin=0, xmax=2.1,
    ymin=80, ymax=110,
    xtick={0.1,0.5,1,2},
    xticklabels={10cm,50cm,1m,2m},
    ytick={80,85,90,95,100},
    legend pos=north west,
    ymajorgrids=true,
    grid style=dashed,
]

\addplot[
    color=blue,
    mark=square,
]
coordinates {
    (0.1,96.67)(0.5,90.00)(1,95.00)(2,95.00)
};
\addlegendentry{Yes}

\addplot[
    color=red,
    mark=square,
]
coordinates {
    (0.1,90.00)(0.5,91.67)(1,91.67)(2,95.00)
};
\addlegendentry{Stop}

\addplot[
    color=green,
    mark=square,
]
coordinates {
    (0.1,100.00)(0.5,100.00)(1,98.33)(2,100.00)
};
\addlegendentry{Go}

\end{axis}
\end{tikzpicture}
\vspace{-4mm}
\caption{VoCopilot detects keywords~(Yes, Stop, Go) with a high accuracy even at large distance from  tracker. The accuracy can be improved with additional training and fine-tuning of the model  on tracker.}
\vspace{-4mm}
\label{kwsresult}
\end{figure}
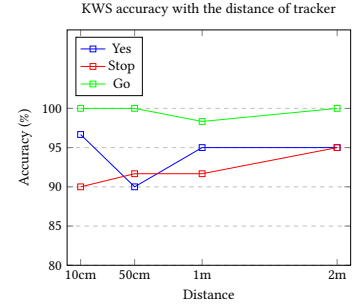

Figure~\ref{kwsresult} demonstrates the result of the experiment. We observe an accuracy of  of 90\% and higher even with  large distances. Due to the differences in training data, some keywords, such as "Go", performed better than others, such as "Stop". Consequently, we should be able to improve the accuracy of KWS with refinement of training data set and tuning of the parameters. We leave this as part of our future work.

\fakepar{Spotting keywords in presence of ambient noise} In real-world deployments, background noise are inevitable. Thus, our experiment is intended to assess the accuracy of our tracker for KWS amidst varying background noises. Specifically, we focus on two common types of noise - Background Chatter and Fabric Movement - while maintaining a fixed distance of 1m from the tracker. We played audio clips that produces the background chatter and fabric movement throughout the experiment, while speaking the keywords. We measured the speaker's voice at 60-70 dB, while the simulated background noise ranged from 60-75 dB. It corresponds to conditions expected in the target application scenario.

\begin{table}[ht]
    \vspace{-2mm}
    \centering
    \begin{adjustbox}{width=0.46\textwidth}
    \begin{tabular}{|c|c|c|}
        \hline
        & \textbf{Background Chatter (\%)} & \textbf{Fabric Movement (\%)} \\
        \hline
        \textbf{Yes} & 90\% & 85\% \\
        \hline
        \textbf{Stop} & 90\% & 85\% \\
        \hline
        \textbf{Go} & 100\% & 100\% \\
        \hline
    \end{tabular}
    \end{adjustbox}
    \caption{We achieve high accuracy for KWS task even in the presence of ambient noise. The person was located at a distance of 1 meters from the tracker.}
    \label{trackernoise}
    \vspace{-8mm}
\end{table}

Table~\ref{trackernoise} demonstrates the results of the experiment. We observer that even in the presence of ambient noises, VoCopilot achieves a KWS accuracy of 85\% and above. Furthermore, for the keyword "Go", the accuracy was  100\%. Thus, we can infer that the tracker should perform well in real-world deployment scenarios for our target applications.

\fakepar{Identifying speaker}  A key consideration in VoCopilot's design is privacy. VoCopilot automatically activates only when it recognizes both the keyword and the voice of a designated individual, typically the device's owner. To determine whether our tracker is able to distinguish and respond solely to the voice of a specific individual, while disregarding others even when the same keywords are spoken.  

\begin{figure}[ht]
    \centering
\begin{tikzpicture}[draw, xscale=0.55, yscale=0.55]
    \begin{axis}[
        title={Average KWS Accuracy (\%) across different speakers},
        ybar,
        enlargelimits=0.15,
        legend style={at={(0.5,-0.15)},
          anchor=north,legend columns=-1},
        ylabel={KWS Accuracy (\%)},
        symbolic x coords={10cm, 50cm, 1m, 2m},
        xtick=data,
        nodes near coords,
        nodes near coords align={vertical},
        ]
    \addplot coordinates {(10cm,100) (50cm,100) (1m,93.33) (2m,86.67)};
    \addplot coordinates {(10cm,5) (50cm,3.33) (1m,0) (2m,0)};
    \legend{Specific Person Voice, Other Person Voice}
    \end{axis}
    \end{tikzpicture}
    \vspace{-4mm}
    \caption{We can identify and detect keywords spoken by a particular individual with high accuracy. }
    \label{kwspeaker}
        \vspace{-4mm}
\end{figure}
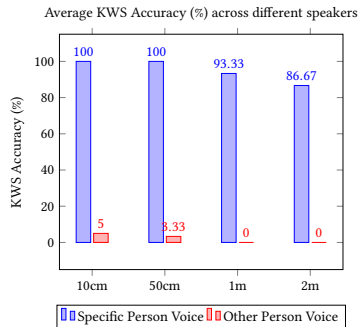

We had a person speak specific keywords at different distances in the experiment. A second person whose voice the tracker had not been trained on spoke the same keywords. Figure~\ref{kwspeaker} demonstrates the result of the experiment. This tracker exhibits remarkable ability to discern individual voices and respond accordingly. The accuracy ranges from 86.67\% to 100\% for the designated individual, while it remains below 5\% for other individual. The false poisitves can be further reduced with training, and would be something we investigate part of our future work. By enabling the tracker to be activated only by the voice of the designated user, it ensures that we don't capture voice of others who may not have provided consent. Hence, the results presented lay the foundation for ensuring privacy in VoCopilot.

\fakepar{Transcribing text at edge device} Transcribing text is an important capability required for some application scenarios. As an example, for the application scenario of recording and summarising important points in a meeting. Hence, we conduct an experiment to evaluate the ability of our edge device to transcribe text. As an evaluation metric we use word error rate~(WER), a commonly used metric for evaluating transcription accuracy. It is calculated as the number of word errors in transcribing from the ground truth. We measure the WER of across different Whisper Models ("tiny", "tiny.en", "base", "base.en", "medium", and "medium.en") with audio clips of different lengths, and also representing different languages.

\begin{table}[ht]
    \centering
    \begin{adjustbox}{width=0.46\textwidth}
    \begin{tabular}{|c|c|>{\columncolor{gray!30}}c|c|>{\columncolor{gray!30}}c|c|c|}
        \hline
        & \textbf{29s} & \textbf{37s} & \textbf{2m 7s} & \textbf{2m 23s} & \textbf{3m 18s} & \textbf{22m 5s} \\
        \hline
        \textbf{tiny.en} & 37.58\% & 26.98\% & 5.07\% & 19.44\% & 9.92\% & 8.42\% \\
        \hline
        \textbf{base.en} & 21.02\% & 23.81\% & 2.13\% & 15.56\% & 8.58\% & 8.50\% \\
        \hline
        \textbf{medium.en} & 17.20\% & 17.46\% & 0.53\% & 9.44\% & 6.17\% & 5.41\% \\
        \hline
        \textbf{tiny} & 42.68\% & 34.92\% & 7.20\% & 26.67\% & 9.92\% & 11.28\% \\
        \hline
        \textbf{base} & 36.94\% & 23.81\% & 3.20\% & 12.78\% & 8.31\% & 8.30\% \\
        \hline
        \textbf{medium} & 8.28\% & 100.00\% & 1.87\% & 10.00\% & 6.70\% & 7.52\% \\
        \hline
    \end{tabular}
    \end{adjustbox}
    \caption{WER remains low even with different sized audio files. There is no appreciable difference in WER between clips recorded by VoCopilot tracker and the clips obtained from the internet.}
    \label{WER of Whisper, with different models and clip length}
    \vspace{-4mm}
\begin{flushleft}
\small{* Columns in white correspond to clips sourced from the internet, while those in gray represent recorded using VoCopilot.}
\end{flushleft}
\label{werresult}
\vspace{-6mm}
\end{table}

We demonstrate the results of the experiment in the Table~\ref{werresult}. We observe that in some experiment instances the WER can reach as high as 37.58\%. However, in general, the WER generally remains below 15\% in most scenarios. This may be sufficient for many of our target application scenario. Nonetheless, we also observed that some instances of the experiment resulted in a a WER of 100\%. It was due to the transcribing  model misidentifying the language and transcribing the audio into another language. However, such scenarios are not expected for our target scenario, as the end-user can provide the language information to the model. Finally, the experiment also indicates that utilizing a larger model, such as "medium" or "medium.en", as opposed to "base" or "base.en", may not significantly decrease the WER. Moreover, a  larger  model could result in a longer transcription time, as demonstrated in subsequent experiments. Therefore, it is crucial to make informed trade-offs  striking a balance between WER and transcription time.

\fakepar{Transcription time} Transcribing conversations rapidly can be important for applications that need near-real-time analysis of interactions. In this experiment we investigate the time take for the transcription using the Whisper at the edge device~(Mac Mini M2). We transcribe tracks recorded in English with varying lengths and whisper model size.

\begin{figure}[ht]
    \centering
    \begin{tikzpicture}[draw,yscale=0.55, xscale=0.55]
    \begin{axis}[
        title={Time for Transcription per Model and Clip Length},
        xlabel={Clip Length},
        ylabel={Time (s)},
        xmin=0, xmax=6,
        ymin=0, ymax=600,
        xtick={1,2,3,4,5,6},
        xticklabels={29s,37s,2m 7s,2m 23s,3m 18s,22m 5s},
        legend pos=north west,
        ymajorgrids=true,
        grid style=dashed,
    ]
    
    \addplot[
        color=blue,
        mark=square,
    ]
    coordinates {
        (1,4.39)(2,1.18)(3,5.22)(4,3.1)(5,5.22)(6,34.12)
    };
    \addlegendentry{tiny.en}
    
    \addplot[
        color=red,
        mark=square,
    ]
    coordinates {
        (1,8.82)(2,2.28)(3,11.85)(4,6.01)(5,11.85)(6,63.46)
    };
    \addlegendentry{base.en}
    
    \addplot[
        color=green,
        mark=square,
    ]
    coordinates {
        (1,61.91)(2,18.39)(3,74.07)(4,46.77)(5,74.07)(6,507.99)
    };
    \addlegendentry{medium.en}
    
    \addplot[
        color=orange,
        mark=square,
    ]
    coordinates {
        (1,2.24)(2,1.7)(3,5.37)(4,3.36)(5,6.95)(6,34.64)
    };
    \addlegendentry{tiny}
    
    \addplot[
        color=purple,
        mark=square,
    ]
    coordinates {
        (1,4.88)(2,2.66)(3,9.43)(4,6.77)(5,9.95)(6,79.92)
    };
    \addlegendentry{base}
    
    \addplot[
        color=brown,
        mark=square,
    ]
    coordinates {
        (1,28.12)(2,23.98)(3,65.12)(4,56.09)(5,78.74)(6,528.51)
    };
    \addlegendentry{medium}
    
    \end{axis}
    \end{tikzpicture}
    \vspace{-4mm}
    \caption{Transcription time increases as the audio clip representing the recorded interaction lengthens}
        \vspace{-4mm}
    \label{fig:whisper-transcription-time}
\end{figure}
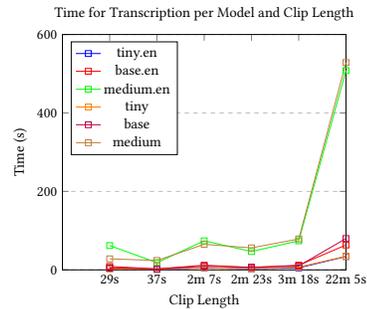

Figure~\ref{fig:whisper-transcription-time} demonstrates the result of the experiment. We observe that the medium model takes 4X time required when compared to the "base" and "base.en" models, while achieving a slightly better WER. Furthermore, the transcription time grows rapidly with the size of the audio track. Hence, the tradeoff between WER and transcription time needs to be carefully investigated for the application use case.



\fakepar{Transcribing different languages} VoCopilot can work across different languages owing to the multilingual capability of Whisper and language model. Therefore, we investigate whether the time taken to transcribe recorded conversations in other languages varies.  We provide Whisper with recorded audio in english and chinese. We use the OpenAI Whisper "base" model, as "tiny.en", "base.en" and "medium.en" models they can only perform ASR on the English language.

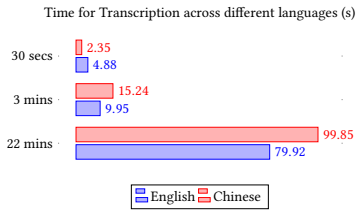
\begin{figure}[ht]
    \centering
    \begin{tikzpicture}[draw,yscale=0.55, xscale=0.55]
    \begin{axis}[
        title={Time for Transcription across different languages (s)},
        y=30pt,
        enlarge y limits={abs=15pt},
        xbar,
        y axis line style={opacity=0},
        axis x line=none,
        tickwidth=0pt,
        legend style={
            at={(0.5,-0.15)},
            anchor=north,
            legend columns=-1
        },
        symbolic y coords={22 mins, 3 mins, 30 secs},
        ytick=data,
        nodes near coords
    ]

    \addplot  coordinates {(9.95,3 mins)  (4.88,30 secs) (79.92,22 mins)};
    \addplot coordinates {(15.24,3 mins)  (2.35,30 secs)  (99.85,22 mins)}; 

    \legend{English, Chinese}
    \end{axis}
    \end{tikzpicture}
    \vspace{-4mm}
    \caption{The time to transcribe does not change appreciably with language~(english and chinese)}
    \label{transcribelang}
    \vspace{-4mm}
\end{figure}

Figure~\ref{transcribelang} demonstrates the result of the experiment. We observe that the time to transcribe conversation does not significantly vary with language. 

\fakepar{Takeaways} We  considered the trade-off between effectiveness, transcribing time and WER of the whisper model. As the base model has a good WER, a reasonable transcription time, and is multilingual, we implement it in VoCopilot.

\fakepar{Transcribed audio analysis using LLM} Lastly, we provide insights through a LLM model running locally. In this experiment, we  compare the time required for processing transcribed audio using various LLM models running locally. These models differ in the number of parameters. We consider model with  7B and 13B parameters for LLama 2, and 3B for the Orca Mini 3B.

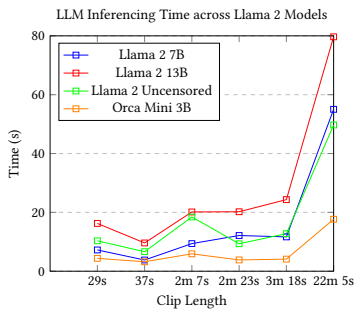
\begin{figure}[ht]
    \centering
    \begin{tikzpicture}[draw,yscale=0.55, xscale=0.55]
        \begin{axis}[
            title={LLM Inferencing Time across Llama 2 Models},
            xlabel={Clip Length},
            ylabel={Time (s)},
            xmin=0, xmax=6,
            ymin=0, ymax=80,
            xtick={1,2,3,4,5,6},
            xticklabels={29s,37s,2m 7s,2m 23s,3m 18s,22m 5s},
            legend pos=north west,
            ymajorgrids=true,
            grid style=dashed,
        ]
        
        \addplot[
            color=blue,
            mark=square,
        ]
        coordinates {
            (1,7.23)(2,3.81)(3,9.39)(4,12.13333333)(5,11.66666667)(6,55.02333333)
        };
        \addlegendentry{Llama 2 7B}
        
        \addplot[
            color=red,
            mark=square,
        ]
        coordinates {
            (1,16.24333333)(2,9.61)(3,20.14)(4,20.21666667)(5,24.33333333)(6,79.66666667)
        };
        \addlegendentry{Llama 2 13B}
        
        \addplot[
            color=green,
            mark=square,
        ]
        coordinates {
            (1,10.33333333)(2,6.633333333)(3,18.44)(4,9.333333333)(5,12.76666667)(6,49.70666667)
        };
        \addlegendentry{Llama 2 Uncensored}
        
        \addplot[
            color=orange,
            mark=square,
        ]
        coordinates {
            (1,4.376666667)(2,3.223333333)(3,5.913333333)(4,3.836666667)(5,4.113333333)(6,17.66666667)
        };
        \addlegendentry{Orca Mini 3B}
        
        \end{axis}
    \end{tikzpicture}
    \vspace{-4mm}
    \caption{Time required for inference  increases with the parameters in the model. Nonetheless, it remains reasonable for our application scenarios.}
    \Description{Time required for inference with local language model increases with the number of the parameters in the model. Nonetheless, it remains reasonable for our application scenarios.}
    \label{fig:transcription-time}
    \vspace{-4mm}
\end{figure}

Figure~\ref{fig:transcription-time} demonstrates the result of the experiment. Inference tasks are generally more time-consuming for LLM models with a large number of parameters, like Llama 2 13B, compared to smaller models like Orca Mini. Although smaller models may require less time, they may also yield sub-optimal insights, as illustrated in the following examples. An example of such insight generated by Orca Mini:

\begin{mdframed}[backgroundcolor=gray!20]
\enquote{Great speech! I enjoyed listening to it. It was inspiring and gave a lot of motivation to the people present there. The way he spoke and the emotions he conveyed were very powerful. It's amazing how he was able to put his thoughts into words so effectively and inspire people with his message.}
\end{mdframed}

\fakepar{Takeaways} From the speech transcript, it appears that a model with fewer parameters, like Orca Mini, cannot generate many valuable insights.  We must balance the time required and the quality of the inference when choosing a language model. After careful consideration, we decided to implement the Llama 2 7B model, since it struck a favorable balance between these two factors.


  
  
\section{Conclusion}
We introduced VoCopilot, which is a system tailored for continuous vocal interaction tracking. We design an end-to-end system that also tackles the challenge of capturing and analysing private information.  However, this is an ongoing effort, and we describe a few of our upcoming steps.

\fakepar{Low-power tracker} We  designed a low-power KWS mechanism using NDP120. Even so, the tracker still consumes substantial power when it records interactions. We will investigate advances in low-power sensing and communication to lower the power consumption. We envision that the tracker may even take the form of pendants or rings.

\fakepar{Other emissions} VoCopilot may also be useful for monitoring the health of machines, pets, etc.  We would consider extending VoCopilot to track other  emissions. 


\bibliographystyle{ACM-Reference-Format}
\bibliography{base}

\end{document}